\documentclass[aps,prl,superscriptaddress,reprint,longbibliography,twocolumn]{revtex4-2}

\usepackage{amsfonts}
\usepackage{subfigure}
\usepackage{amsmath}
\usepackage{amssymb}
\usepackage{amsbsy} 
\usepackage{bbm}
\usepackage{epsfig}
\usepackage{graphicx}
\usepackage{epstopdf}
\usepackage{color, xcolor}
\usepackage{mathdots}
\usepackage{braket}
\usepackage{latexsym}
\usepackage{amsthm}
\usepackage{hyperref}

\hypersetup{hypertex=true, colorlinks=true, linkcolor=blue, anchorcolor=blue, citecolor=blue,urlcolor=blue}

\renewcommand{\Re}{\operatorname{Re}}
\renewcommand{\Im}{\operatorname{Im}}

\def\be{\begin{equation}} \def\ee{\end{equation}}
\def\bea{\begin{eqnarray}} \def\eea{\end{eqnarray}}

\newcommand{\id}{\mathbbm{1}}

\newtheorem{lemma}{Lemma} 
\newtheorem{definition}{Definition} 

\begin{document}

	\title{Geometric Origin of Non-Bloch PT Symmetry Breaking}
	\author{Yu-Min Hu}
	\affiliation{Institute for Advanced Study, Tsinghua University, Beijing, 100084, China}
	\author{Hong-Yi Wang}
	\affiliation{Institute for Advanced Study, Tsinghua University, Beijing, 100084, China}
	
	\author{Zhong Wang}
	\affiliation{Institute for Advanced Study, Tsinghua University, Beijing, 100084, China}
	
	\author{Fei Song}
	\altaffiliation{songfeiphys@gmail.com}
	\affiliation{Institute for Advanced Study, Tsinghua University, Beijing, 100084, China}
	\affiliation{Kavli Institute for Theoretical Sciences, Chinese Academy of Sciences, Beijing, 100190, China }
	
	\begin{abstract}
		The parity-time (PT) symmetry of a non-Hermitian Hamiltonian leads to real (complex) energy spectrum when the non-Hermiticity is below (above) a threshold. Recently, it has been demonstrated that the non-Hermitian skin effect generates a new type of PT symmetry, dubbed the non-Bloch PT symmetry, featuring unique properties such as high sensitivity to the boundary condition. Despite its relevance to a wide range of non-Hermitian lattice systems, a general theory is still lacking for this generic phenomenon even in one spatial dimension. Here, we uncover the geometric mechanism of non-Bloch PT symmetry and its breaking. We find that non-Bloch PT symmetry breaking occurs by the formation of cusps in the generalized Brillouin zone (GBZ). Based on this geometric understanding, we propose an exact formula that efficiently determines the breaking threshold. Moreover, we predict a new type of spectral singularities associated with the symmetry breaking, dubbed non-Bloch van Hove singularity, whose physical mechanism fundamentally differs from their Hermitian counterparts. This singularity is experimentally observable in linear responses.
	\end{abstract}
	
	\maketitle
	
	\emph{Introduction.--} Parity-time (PT) symmetry is one of the central concepts in non-Hermitian physics \cite{bender1998real,Bender2002,bender2007making,Mostafazadeh2002i, Mostafazadeh2002ii}. A PT-symmetric Hamiltonian enjoys a real-valued spectrum when the non-Hermiticity is below a certain threshold. Above this threshold, the symmetry-protected reality breaks down. This real-to-complex transition has been associated with the exceptional points (EP) where a pair of eigenstates coalesce \cite{Ozdemir2019review,el2018non,Miri2019review,heiss2004exceptional,berry2004physics}. The unique properties of PT symmetry and EP have inspired numerous explorations on various experimental platforms \cite{makris2008beam, guo2009complex, lin2011unidirectional,regensburger2012parity, bittner2012, hodaei2014PT, feng2014singlemode, fleury2015invisible, hodaei2017enhanced}.

	Recently, the non-Hermitian skin effect (NHSE) has been realized as a general mechanism for achieving PT symmetry and therefore real spectrums \cite{Longhi2019Probing,Longhi2019nonBloch,xiao2021observation, weidemann2022topological,yang2022designing, zhang2022real, zeng2022real}.  NHSE refers to the phenomenon that the eigenstates of non-Hermitian systems are squeezed to the boundary under open boundary condition (OBC), which causes strong sensitivity of the spectrum to boundary conditions \cite{yao2018edge,yao2018chern,kunst2018biorthogonal,
		lee2018anatomy,alvarez2017,Helbig2019NHSE,Xiao2019NHSE, Ghatak2019NHSE,Weidemann2020, Hofmann2020,Bergholtz2021RMP,Ashida2020}. Its quantitative description requires a non-Bloch band theory that generalizes the concept of Brillouin zone \cite{yao2018edge,Yokomizo2019,Yokomizo2020review,sczhangmemorial,
		Yang2019Auxiliary,Longhi2020chiral,Kawabata2020nonBloch,Deng2019}. In the presence of NHSE, it is possible to have an entirely real spectrum under OBC, in sharp contrast to that under periodic boundary condition (PBC), which is always complex \cite{Zhang2020correspondence,Okuma2020}. That the real spectrum can only be maintained under OBC is known as \emph{non-Bloch PT symmetry} \cite{Longhi2019Probing,Longhi2019nonBloch,xiao2021observation}, which is crucial in the experimental detection of non-Bloch band topology \cite{Xiao2019NHSE,wang2021detecting}. Recent experiments have confirmed the non-Bloch PT symmetry breaking transitions, i.e., the real-to-complex transitions of the OBC spectrum \cite{xiao2021observation, weidemann2022topological}.

	The PT symmetry breaking in Bloch bands originates exclusively from the Bloch Hamiltonian being defective at certain wave vectors \cite{makris2008beam,regensburger2012parity,longhi2009bloch,bender2015wave,stegmaier2021topological}. The non-Bloch PT symmetry breaking, however, must have an entirely different mechanism. One of the clear evidences is that the non-Bloch PT breaking can occur in single-band systems. In contrast, the Bloch PT breaking is strictly prohibited in a single-band system because its Bloch Hamiltonian, as a complex number, can never be defective. It is the purpose of this paper to unveil the mechanism of non-Bloch  PT breaking.

	We uncover a geometric origin of non-Bloch PT symmetry breaking and formulate a coherent theory that enables efficient computation of the PT breaking threshold in one dimension (1D). Specifically, the geometric object we will focus on is the generalized Brillouin zone (GBZ), which can be determined through 1D non-Bloch band theory \cite{yao2018edge,Yokomizo2019}. Because of its noncircular shape, the GBZ can possibly have intriguing cusp singularities \cite{Yokomizo2019,Yokomizo2020topological}, yet their physical significance remains elusive. Our work starts from the observation that these singularities underlie the non-Bloch PT symmetry breaking.  Our main results include: (i) The cusps on a GBZ are responsible for the non-Bloch PT symmetry breaking. (ii) A concise formula is found for the PT breaking threshold that does not require calculating the energy spectrum or GBZ. (iii) The transition point of non-Bloch PT symmetry breaking represents a new type of divergence in the density of states (DOS), which we call the \emph{non-Bloch van Hove singularity}. 
	
\begin{figure}
		\includegraphics[width=8cm]{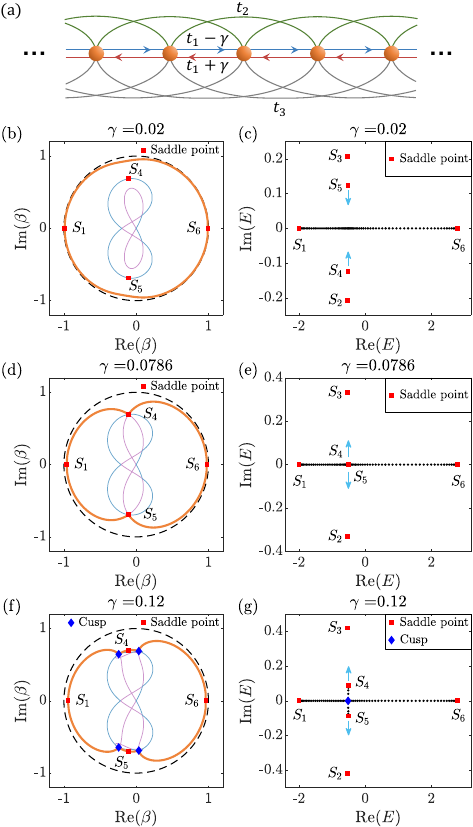}
		\caption{(a) The real-space hopping in the model of Eq.~(\ref{model}). (b)-(g) The transition of the GBZ and the energy spectrum in a representative breaking process, with parameters $t_1=1, t_2=t_3=0.2$. The energy spectrums are obtained by diagonalizing the real-space Hamiltonian under OBC with $L=80$. (b),(c) The PT-exact phase with $\gamma=0.02$; (d),(e) the transition point with $\gamma = 0.0786$; and (f),(g) The PT-broken phase with $\gamma=0.12$. In (b),(d),(f), the black dashed loop is the BZ, the orange loop is the GBZ, and other branches in the aGBZ are labeled with different colors. $S_{2}$ and $S_{3}$ in (c),(e),(g) are outside the plot region of (b),(d),(f).}
		\label{fig1}
	\end{figure}

	\emph{Geometric origin.--} Non-Bloch PT symmetry breaking refers to the real-to-complex transition of the OBC spectrum in non-Hermitian bands. We first attempt to gain some intuitions about this transition from a concrete example. A simple model that includes all the ingredients of our interest has the Bloch Hamiltonian:\bea
	H(k)=2t_1\cos k+2t_2\cos2k+2t_3\cos3k+2i\gamma\sin k\label{model}.
	\eea
	Its real-space hopping is illustrated in Fig.~\ref{fig1} (a). The real-space Hamiltonian $H$ has real matrix elements and hence obeys the (generalized) PT symmetry $\mathcal{K} H \mathcal{K}=H$, where $\mathcal{K}$ is the complex conjugate operator \cite{Bender2002}. The (generalized) PT symmetry is essential for obtaining a robust non-Bloch PT-exact phase (i.e., a parameter region with nonzero measure where OBC spectrums are purely real).
	
	The standard approach to obtain the OBC spectrum of the model in Eq.~(\ref{model}) is to use the non-Bloch band theory, which takes into account the NHSE \cite{yao2018edge,Yokomizo2019}. In this approach, the Bloch Hamiltonian is generalized to the complex plane $H(\beta) \equiv H(k)|_{e^{ik}\to\beta}$, dubbed non-Bloch Hamiltonian. The OBC spectrum is given by $H(\beta)$, where $\beta$ is taken from the GBZ, rather than the Brillouin zone (BZ). The GBZ is a curve determined by the GBZ equation $|\beta_i(E)| = |\beta_{i+1}(E)|$, where $\beta_i(E)$ and $\beta_{i+1}(E)$ are the middle two among all roots of the characteristic function $f(E,\beta)=\text{det}[H(\beta)-E \id]=0$ sorted by their moduli \footnote{Sorting all roots by their moduli, $\beta_i(E)$ and $\beta_{i+1}(E)$ are the $i$th and $(i+1)$-th roots, respectively, where $i$ is the order of the pole of the Laurent polynomial $f(E,\beta)$ at $\beta=0$. For example, the model Eq.~(\ref{model}) with nonzero $t_3$ has the index $i=3$. }. Thus, the decay factor (also known as the inverse skin depth) of a non-Hermitian skin mode is given by $\ln |\beta|$ with $\beta\in\text{GBZ}$. Numerically, an efficient approach to solve the GBZ is to first obtain the so-called auxiliary GBZ (aGBZ) \cite{Yang2019Auxiliary}, which comprises a bunch of curves satisfying $|\beta_i(E)|=|\beta_j(E)|$ for any $i \neq j$. Then the GBZ comes as a subset of the aGBZ by further choosing the indices of these roots with equal moduli.
	
	In Fig.~\ref{fig1}, we demonstrate a paradigmatic non-Bloch PT transition within the model Eq.~(\ref{model}). With increasing $\gamma$, the OBC spectrum changes from entirely real to partially complex. Moreover, the continuity of GBZ changes saliently before and after the transition point. The GBZ is completely smooth in the PT-exact phase [Fig.~\ref{fig1}(b)], but becomes singular at several cusps when PT symmetry is broken [Fig.~\ref{fig1}(f)]. Remarkably, the cusps appear exactly at the transition point [Fig.~\ref{fig1}(d)].

	At the same time, we mark \emph{saddle points} that satisfy $f(E,\beta)=\partial_\beta f(E,\beta)=0$ \cite{Longhi2019Probing} by the red points in Fig.~\ref{fig1}. Tracking their motions can help us understand the generation of GBZ cusps. A saddle point must reside on the aGBZ, but it may or may not be on the GBZ \cite{Yang2019Auxiliary, footnote1}.  In Fig.~\ref{fig1}(b), $S_4$ and $S_5$ reside on the aGBZ but not on the GBZ. However, as indicated in Fig.~\ref{fig1}(d) and Fig.~\ref{fig1}(f), they are merged into the GBZ at the transition point. For this to be possible, at the transition point the GBZ intersects with multiple branches of the aGBZ [Fig.~\ref{fig1}(d)], which results in $S_4$ and $S_5$ being saddle points and GBZ cusps simultaneously. 
	
	To interpret the above observations, we parameterize the GBZ as $\beta= |\beta(\theta)| e^{i\theta}$. It suffices that a parametrization exists in a neighborhood of a given $\beta$. The derivative of the energy dispersion $E(\theta)=H(|\beta(\theta)|e^{i\theta})$ with respect to angle $\theta$ is
	\bea
	\frac {d E(\theta)}{d \theta}=\frac {\partial H(\beta)}{\partial \beta}\left(\frac{\partial |\beta(\theta)|}{\partial \theta}e^{i\theta}+i\beta\right).\label{geometric origin}
	\eea
	The cusps correspond to discontinuous points of $\partial|\beta(\theta)|/\partial \theta$, and thus $d E(\theta)/d\theta$ is also discontinuous at the cusp unless $\partial_\beta H(\beta)=0$. It is this discontinuity that accounts for the multiple branches of the spectrum on the complex plane, and the branch point is just the cusp energy. This explains why the spectrum in the PT-exact phase simply lies on the real axis [Fig.~\ref{fig1}(c)], while it becomes complex and ramified in the PT-broken phase [Fig.~\ref{fig1}(g)]. In the critical cases [Figs.~\ref{fig1}(d)-(e)], a cusp appears but the spectrum is still entirely real. This is only possible when these cusps are also saddle points satisfying $\partial_\beta H(\beta)=0$ \footnote{The reality of the spectrum guarantees $d_\theta\Im E(\theta)=0$. Meanwhile, one can derive $d_\theta\Im E(\theta)=\Re[\beta \partial_\beta H(\beta)]+ \Im[\beta \partial_\beta H(\beta)] \partial_\theta \log |\beta(\theta)|$ from Eq.~(\ref{geometric origin}). For a cusp $\beta_c=|\beta(\theta_c)|e^{i\theta_c}\in \text{GBZ}$ where $\left(\partial|\beta(\theta)|/\partial \theta\right)_{\theta\rightarrow \theta_c^-}\neq \left(\partial|\beta(\theta)|/\partial \theta\right)_{\theta\rightarrow \theta_c^+}$, it follows that $\Re[\beta \partial_\beta H(\beta)]=\Im[\beta \partial_\beta H(\beta)]=0$. Since GBZ does not pass $\beta=0$ in general, we conclude that $\beta_c$ should be a saddle point.}.

	The above analysis indicates that the emergence of GBZ cusps is a geometric origin of non-Bloch PT symmetry breaking. The Supplemental Material \cite{footnote1} includes more examples that demonstrate this cusp mechanism, which can be generally formulated as follows:

	\begin{itemize}
		\item[(i)] A PT-symmetric lattice system has a smooth GBZ if it is in the non-Bloch PT-exact phase;
		\item[(ii)] If there are cusps on the GBZ, the system is either in the non-Bloch PT-broken phase or at the PT transition point.
	\end{itemize}
	
	The proof of this result is given in \cite{footnote1}. It leverages a basic property of arbitrary non-Hermitian lattice systems with short-range hoppings, namely, the analyticity of the characteristic polynomial $f(E, \beta)$ with respect to $\beta$ and $E$.

	\emph{Simple formula for the breaking threshold.--} In addition to the geometric origin, another piece of information conveyed by the model Eq.~(\ref{model}) is that its PT transition is characterized by the motion of saddle-point energies. With increasing non-Hermiticity, the energies of $S_4$ and $S_5$ move upward and downward, respectively [Figs.~\ref{fig1}(c),(e),(g)]. Notably, along with $S_4$ and $S_5$ being merged into the GBZ [Fig.~\ref{fig1}(d)], their energies coalesce on the real axis at the transition point [Fig.~\ref{fig1}(e)]. For a single-band model with non-Bloch Hamiltonian $H(\beta)=\sum_{n=-l}^{r} h_n \beta^n$, such a coalescence is described by 
	\begin{equation}    \label{coalescence}
		H(\beta_{s,i}) = H(\beta_{s,j}) \in \mathbb{R},
	\end{equation}
	where $\beta_{s,i}$ and $\beta_{s,j}$ are two different saddle points on the GBZ, satisfying $ \partial_\beta H(\beta)=0$. We will demonstrate that the condition Eq.~(\ref{coalescence}) serves as an efficient criterion for determining the non-Bloch PT breaking threshold. 
	
	We shall rephrase Eq.~(\ref{coalescence}) in two steps to make its identification more feasible. First, we utilize a mathematical concept called \emph{resultant} to search for any degeneracy of saddle-point energies, i.e., $H(\beta_{s,i})=H(\beta_{s,j})$ with $i\neq j$. Then, we locate the parameter values that fulfill the condition Eq.~(\ref{coalescence}) by examining both the reality of the degenerate energies and whether the associated saddle points belong to the GBZ. Here, the resultant is defined to identify whether two given polynomials have a common root \cite{lang2002algebra}. For example, the resultant $\operatorname{Res}_x[x-a,x-b]=a-b$ equals zero if and only if the roots $x=a$ and $x=b$ are degenerate. Recalling that the saddle points are exactly the common roots of $f(E,\beta)=H(\beta)-E=0$ and $\partial_\beta f(E,\beta)=\partial_\beta H(\beta)=0$, saddle-point energies $E_{s,i}=H(\beta_{s,i})$ can be directly found by eliminating $\beta$, which results in
	\bea    \label{resultant1}
	g(E) = \operatorname{Res}_\beta \left[ \tilde{f}(E,\beta),\partial_\beta \tilde{f}(E,\beta) \right]=0,
	\eea
	where $\tilde{f}(E,\beta)=\beta^l f(E,\beta)$ is used to avoid negative powers of $\beta$. The roots of $g(E)=0$ are exactly all the saddle-point energies $E_{s,i}$, i.e., $g(E) \propto \prod_i (E-E_{s,i})$. On the other hand,  the coalescence condition Eq.~(\ref{coalescence}) suggests at least a pair of $E_{s,i}$ are degenerate, which is thus equivalent to $\partial_E g(E)=0$. Therefore, the parameters with degenerate saddle-point energies can be solved from
	\bea \label{resultant2}
	\operatorname{Res}_E \left[ g(E),\partial_E g(E) \right] = 0.
	\eea
	A standard procedure to derive the above resultants is through the Sylvester matrix \cite{footnote1}.
	
	When we consider $\gamma$ variable and other parameters fixed, $\operatorname{Res}_E[g(E),\partial_E g(E)]$ is nothing but a polynomial of $\gamma$. We are now in a place to tell which root of this polynomial truly contributes to the coalescence described by Eq.~(\ref{coalescence}). In practice, the desired root is recognized under the following procedure. We insert $\gamma$ obtained from Eq.~(\ref{resultant2}) back into Eq.~(\ref{resultant1}) to find out the degenerate energies $E_s$. Then, we solve and sort the roots of $f(E_s,\beta)=\sum_{n=-l}^{r} h_n \beta^n-E_s=0$ as $|\beta_1(E_s)|\leq\ldots\leq|\beta_{l+r}(E_s)|$. Moreover, since Eq. (\ref{resultant2}) is equivalent to the existence of a pair of saddle points with the same energy, we can find two roots $\beta_{s,i}$ and $\beta_{s,j}$ from $f(E_s,\beta)=\partial_\beta f(E_s,\beta)=0$. Finally, according to the GBZ equation and Eq.~(\ref{coalescence}), the PT breaking threshold is determined by selecting those roots of Eq.~(\ref{resultant2}) that fulfill the conditions $\Im E_s=0$ and $|\beta_{s,i}|=|\beta_{s,j}|=|\beta_l(E_s)| = |\beta_{l+1}(E_s)|$ \footnote{Numerically, $|\beta|=|\beta'|$ is verified as $\left||\beta|-|\beta'|\right|<\epsilon$, where we take $\epsilon$ to be $10^{-5}$.}.

	\begin{figure}
		\centering
		\includegraphics[width=8cm]{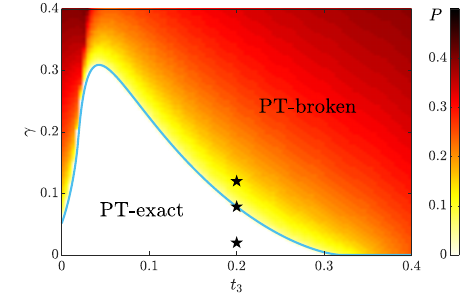}
		\caption{The non-Bloch PT phase diagram of the model Eq.~(\ref{model}) with $t_1=1,t_2=0.2$. The blue line is the phase boundary determined by solving Eq.~(\ref{resultant2}). The color map is a density plot for the proportion $P$ of complex eigenvalues, obtained by counting the proportion of eigenenergies with $|\Im E| > 10^{-10}$. The eigenenergies are obtained by diagonalizing an OBC Hamiltonian of length $L=200$. The three black stars mark the parameters used in Fig.~\ref{fig1}.}
		\label{phasediagram}
	\end{figure}
	So far, we have built up a systematic algebraic method for determining the breaking threshold, the power of which lies in the fact that we are able to find the phase boundary without diagonalizing the real-space Hamiltonian or calculating the complete GBZ. In the Supplemental Material \cite{footnote1}, we explicitly illustrate how to conduct this method step by step for the model Eq.~(\ref{model}) with $t_2=t_3=0$. For more general parameters ($t_2$, $t_3$ nonzero), filtering the roots of Eq.~(\ref{resultant2}) with the GBZ equation is also accurate and effortless for determining the phase boundary. We demonstrate its results in Fig.~\ref{phasediagram}: the boundary between PT-exact and PT-broken phases in the model Eq.~(\ref{model}) obtained via diagonalization agrees well with the one through proper selection of the roots of Eq.~(\ref{resultant2}). The analytic method is, however, much more efficient and free of finite-size effects. 
	
	Not limited to single-band cases, on a non-Hermitian chain with PT symmetry, the generation of the first pair of complex conjugate saddle-point energies contained in the OBC spectrum inevitably involves a coalescence like Eq.~(\ref{coalescence}). This process, whose transition point is predicted by the method introduced here, is experimentally detectable in wave-packet dynamics  \cite{Longhi2019Probing,xiao2021observation,weidemann2022topological}.

	\emph{Non-Bloch van Hove singularity.--} As previously mentioned, at the PT transition point [Figs.~\ref{fig1}(d)-(e)], there exists saddle points on the GBZ that are also cusps. This is a hallmark of the geometric origin of PT-breaking transitions. We shall elucidate the observable consequences of these cusps by examining the non-Hermitian Green's function, defined as $G(E)=(E-H)^{-1}$, where $H$ is the OBC Hamiltonian generated by the Bloch Hamiltonian [e.g. Eq.~(\ref{model})]. Practically, $G(E)$ can be measured through frequency-dependent linear responses on various platforms such as topolectrical circuits \cite{Helbig2019NHSE,li2021quantized}, scattering processes\cite{zirnstein2021bulk}, and open quantum systems \cite{xue2021simple}.

	In the PT-exact phase, we define the DOS along the real axis by $\rho(E)=(\pi L)^{-1} \Im \operatorname{Tr}[G(E+i0^+)]$, or, equivalently, $\rho(E)=L^{-1}\sum_{i=1}^L \delta(E-E_i)$, where $E$ and the eigenenergies $E_i$ are all real. When the system size $L$ goes to infinity, the summation over all eigenenergies becomes an integral along GBZ \cite{xue2021simple, footnote1}. Thus, we have
	\bea \label{eq:dos}
	\rho(E)=\frac{1}{2\pi}\sum_{\beta(E)\in \mathrm{GBZ}} \left| \Im \left[\frac{1}{\beta\partial_\beta H(\beta)}\right]_{\beta=\beta(E)}\right|    ,
	\eea
	which is a natural extension of the well-known formula $\rho(E)=\frac{1}{2\pi}\sum_{E(k)=E} |\partial E(k)/\partial k|^{-1}$ for the Hermitian case.

	\begin{figure}
		\centering
		\includegraphics[width=8cm]{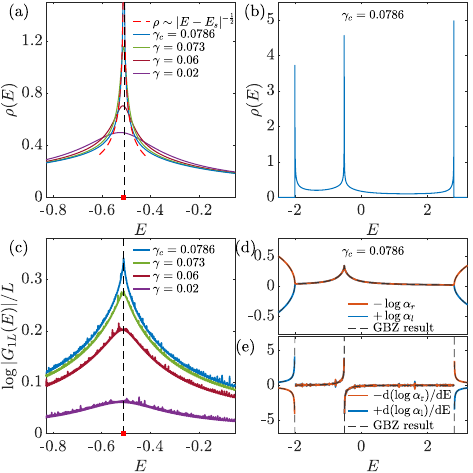}
		\caption{Non-Bloch van Hove singularity. (a) The DOS in the non-Bloch PT-exact phase.  (b) A full profile of DOS at the transition point $\gamma_c=0.0786$. (c)-(d) The frequency dependence of $\log\alpha_{l}(E)=\log|G_{1L}(E)|/L$. (e) $\mathrm{d}(\log\alpha_{l,r}(E))/\mathrm{d}E$ for the scaling factors in (d). The dashed lines in (d)-(e) mark the theoretical predictions based on GBZ. To reduce data fluctuations due to finite-size effects, the $\alpha_{l,r}(E)$ in (d)-(e) are obtained by fitting $\log|G_{1L}(E)|$ and $\log|G_{L1}(E)|$ with respect to the system size $L$.  We fix $L=500$ in (c) and take $L\in [100,300]$ in (d)-(e). The parameters are $t_1=1, t_2=t_3=0.2$. }
		\label{DOS}
	\end{figure}

	According to Eq.~(\ref{eq:dos}), the DOS is divergent at any saddle point on the GBZ. From Figs.~\ref{DOS}(a)-(b), we find that the DOS near $E_s\approx-0.5096$ increases and eventually becomes divergent at the transition point. This divergence is analogous to the van Hove singularity in Hermitian systems, but is induced by the singular shape of the GBZ, which is unique to systems with NHSE.  At the non-Bloch PT symmetry breaking point, the cusps, which are at the same time saddle points, are responsible for the divergence at $E_s$. Thus, we coin for this divergence the term \emph{non-Bloch van Hove singularity}. 
	
	Quantitatively, the asymptotic behavior of the DOS near a non-Bloch van Hove singularity can be inferred from Eq.~(\ref{eq:dos}). Near a saddle point $E_s$, $|\beta\partial_\beta H(\beta)|$ behaves like $|E-E_s|^{\alpha}$. Inserting this back into Eq.~(\ref{eq:dos}), we find that the DOS is locally $\rho(E)\sim |E-E_s|^{-\alpha}$. Generally, the exponent $\alpha$ for a $k$th order saddle point (satisfying $H(\beta)-E=\partial_{\beta} H(\beta)=\ldots=\partial_{\beta}^{k-1} H(\beta)=0$) is $\alpha=1-1/k$. Our model with nonzero $t_3$ gives $\alpha=1/2$, which is in accordance with the numerical fitting $\rho\sim |E-E_s|^{-1/2}$ shown in Fig.~\ref{DOS}(a). Interestingly, in our model with $t_3=0$, two second-order saddle points are merged into one third-order saddle point at the transition point of non-Bloch PT breaking. According to $\alpha=1-1/k$, this also implies that the exponent suddenly changes at the transition point. More details about the jump of $\alpha$ can be found in \cite{footnote1}.
	
	Beyond divergent DOS, the non-Bloch van Hove singularity also manifests itself in the off-diagonal elements of $G(E)$. On a finite chain with length $L$, the end-to-end Green's functions exhibit exponential growth or decay, represented as $|G_{L1}(E)|\sim \alpha_r(E)^L$ and $|G_{1L}(E)|\sim \alpha_l(E)^L$. The two scaling factors $\alpha_r(E)$ and $\alpha_l(E)$ can be predicted using non-Bloch band theory \cite{xue2021simple,hu2023multiband}. For the model Eq.~(\ref{model}) with $t_3\ne0$, we have $\alpha_r(E)=|\beta_3(E)|$ and $\alpha_l(E)=|\beta_{4}(E)|^{-1}$, where $|\beta_{3,4}(E)|$ are the roots of $H(\beta)=E$ sorted as $|\beta_1(E)|\leq \ldots\leq |\beta_6(E)|$. When $E$ belongs to the OBC spectrum, $\alpha_{r,l}(E)$ encode crucial information about GBZ. We find that the frequency dependence of $\alpha_{r,l}(E)$ exhibits a cusp precisely at the energy of the non-Bloch van Hove singularity [Figs.~\ref{DOS}(c)-(d)]. This occurs concurrently with the emergence of GBZ cusps at the transition point.  Furthermore, since these GBZ cusps are also saddle points, the nonsmoothness of $\alpha_{r,l}(E)$ stems from divergent $\mathrm{d}\alpha_{r,l}(E)/\mathrm{d}E$ \footnote{We notice that $\mathrm{d} \alpha_{r,l}(E)/\mathrm{d}E$ remains finite on one side of the left or right band edge. As explained in the Section II of \cite{footnote1}, this is closely related to the fact that the saddle points $S_{1,6}$ shown in Fig.~\ref{fig1}(d), which contribute the energies of two band edges, are not cusps.} [Fig.~\ref{DOS}(e)]. Practically, the divergence in $\mathrm{d} \alpha_{r,l}(E)/\mathrm{d}E$ signals extreme frequency sensitivity in the response to the input signal, which could potentially inspire designs of non-Hermitian sensors \cite{budich2020non, mcdonald2020exponentially}.

	\emph{Conclusions.--} We have presented a theory of non-Bloch PT symmetry breaking in one dimension, which not only explains its geometric origin but also provides an efficient formula for the threshold. Given the fact that the concept of GBZ has recently been generalized to non-Hermitian continuum systems \cite{longhi2021continuum, guo2022continuum, Yokomizo2022continuum,hu2023nonbloch} and disordered systems \cite{ZHANG2023disorder,liu2023modified_GBZ}, it is an interesting direction to develop our theory in these contexts. Moreover, it is known that the threshold in higher dimensions universally approaches zero as the system size increases \cite{song2022non}. In view of the lastest progress on higher-dimensional GBZ \cite{wang2022amoeba,Jiang2023dimensional},  our theory can have implications for PT symmetry in higher dimensions.  For example, since the non-Bloch van Hove singularities are tied to the non-Bloch PT symmetry breaking in one dimension, their proliferation in higher dimensions may be responsible for the universal thresholdless behavior, which is left for future studies. Back to the 1D cases that are experimentally most convenient, our predictions can be verified on various state-of-the-art platforms such as cold atom \cite{li2019observation,Gou2020tunable} and quantum optics systems \cite{xiao2021observation}.

	{\it Acknowledgements.--} This work is supported by NSFC under Grant No. 12125405.
	
	\bibliography{dirac}
	\newpage


	
	
	

	\onecolumngrid

	\section*{The supplemental material}
	\section{Two useful properties of saddle points}
	
	In this section, we prove two useful properties concerning saddle points on the GBZ that have been used in the main text. They are: (i). \emph{A saddle point is a member of the auxiliary generalized Brillouin zone (aGBZ)}. (ii). \emph{The ends of the OBC bulk spectrum always correspond to saddle points}.
	
	To begin, we recapitulate some key definitions. The saddle point in non-Bloch context refers to the solution to both $H(\beta)-E = 0$ and $\partial_\beta\left[H(\beta)-E \right] = 0$, where $H(\beta)$ takes the general form of a non-Bloch Hamiltonian, i.e., $H(\beta)=\sum_{n=-l}^{r} h_n \beta^n$. In the following expressions, we use the subscript ``$s$'' for saddle point values. Based on basic knowledge about algebraic equations, the defining property of the saddle point allows us to make the factorization
	\begin{eqnarray}
		\beta^l\left[H(\beta)-E_s\right]=h_{r}\prod_{i=-l}^{r}(\beta-\beta_i)\propto (\beta-\beta_s)^n\prod_{\beta_i\neq \beta_s}(\beta-\beta_i). \label{denerate}
	\end{eqnarray}
	Here, $n\ge 2$ is the order of the saddle point. Alternatively, we can say that $\beta_s$ is a multiple root of $H(\beta) - E_s = 0$. This reminds us that aGBZ demands that there are two roots of the equation $E=H(\beta)$ sharing the same modulus. Denoting the solutions to this equation as $\beta_i, i=1,2,\dots l+r$, this is to say that there exists a pair $i \neq j$ satisfying
	\begin{equation}    \label{eq:aGBZ}
		\left| \beta_i(E) \right| = \left| \beta_j(E) \right|.
	\end{equation}
	So the condition Eq.~(\ref{eq:aGBZ}) is automatically fulfilled by the multiple roots at a saddle point. Up to now, we have proved the first statement that a saddle point is always a member of the aGBZ.
	
	The second statement directly connects certain key features of the OBC spectrum to saddle points, but we need a little more effort to see this. We start by quoting a theorem stating that the range of the bulk spectrum of a 1D single-band Hamiltonian is geometrically a union of smooth curve segments, joined at a finite number of points \cite{bottcher2005spectral}. In fact, in the framework of non-Bloch band theory, these joining points correspond to the GBZ cusps.  Moreover, the range, as a geometric object on the complex plane, encloses zero area \cite{Okuma2020, Zhang2020correspondence}. Therefore, we can safely talk about its ends without ambiguity. We now parameterize the $\beta$ on GBZ as $\beta(\theta)=|\beta(\theta)|e^{i\theta}$, where $\theta$ stands for the angle of a complex number. In order that the following derivation is well-defined, a parametrization in a neighborhood of the point of concern is sufficient. The bulk state energy depends on $\theta$ through
	\begin{equation}
		E(\theta) = H(\beta(\theta)) = H(|\beta(\theta)| e^{i\theta})   .
	\end{equation}
	Locally, near one end, $E$ smoothly reaches an extremum and then turns back, as $\theta$ varies. Thus, it is easy to see these result in
	\begin{equation}
		\dfrac{dE(\theta)}{d\theta} = 0
	\end{equation}
	evaluated at the turning point $\theta=\theta_{\text{t}}$. More explicitly,
	\begin{equation}
		\dfrac {\partial H(\beta)}{\partial \beta}\left(\dfrac{\partial |\beta(\theta)|}{\partial \theta}e^{i\theta}+i\beta\right) = 0    .
	\end{equation}
	But the bracket above never vanishes, since the first and the second term are perpendicular to each other. So finally we reach
	\begin{equation}
		\left. \partial_\beta H(\beta) \right|_{\beta=\beta_{\text{t}}}= 0 ,
	\end{equation}
	which means an end, namely a turning point of the spectrum, indeed corresponds to a saddle point. However, one should note that the converse is in general not true since saddle points do not necessarily belong to the GBZ, i.e., the corresponding energies do not belong to the OBC bulk spectrum. 
	\section{Proof of cusp mechanism}\label{sec:proof}
	
	In the main text, we have explained the geometric origin of non-Bloch PT symmetry breaking in single-band models based on parameterizing GBZ. We also point out that this type of PT transition is quite general and has its roots in the conformality of analytic functions. In this section, we will complete this argument and prove the following statements: 
	\begin{itemize}
		\item[(i) ] A PT-symmetric lattice system has a smooth GBZ if it is in the non-Bloch PT-exact phase; 
		\item[(ii)] If there are cusps on the GBZ, the system is either in the non-Bloch PT-broken phase or at the PT transition point. 
	\end{itemize}
	
\begin{figure}[h]
		\centering
		\includegraphics[width=14cm]{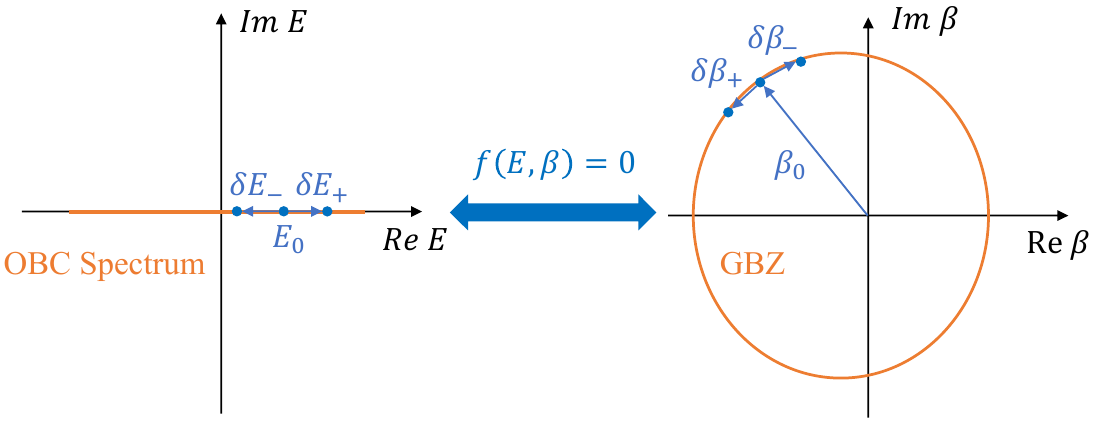}
		\caption{\label{fig:sketch} An auxiliary sketch for proving the key statements at the beginning of Sec. \ref{sec:proof}. }
	\end{figure}

	To elucidate this, we need to specify the precise definitions for the terms ``smooth GBZ" and ``GBZ cusps." GBZ is a collection of one or more curves on the complex plane of $\beta$. If the GBZ is not self-crossing, in the vicinity of a specific point $\beta_0 \in \text{GBZ}$, we can assign a direction to the GBZ.  Then, according to this direction, we take two points, namely $\beta_0+\delta \beta_+ \in \text{GBZ}$ on one side and $\beta_0+\delta \beta_-\in \text{GBZ}$ on the opposite side. As the magnitude of $\delta\beta_\pm$ approaches zero (i.e., $|\delta\beta_{\pm}|\to0$), the vectors $(\Re \delta \beta_+, \Im\delta \beta_+)$ at $\beta_0^+$ and $(\Re \delta \beta_-, \Im\delta \beta_-)$ at $\beta_0^-$ become aligned with the tangents of the GBZ at the points $\beta_0^+$ and $\beta_0^-$, respectively, where the points $\beta_0^{\pm}$ are infinitely close to $\beta_0$ from different directions along the GBZ. (To better understand this point, one may look at Fig.~\ref{fig:sketch}.) We define a GBZ to be smooth when each point on it has a unique tangent line. From this viewpoint, a GBZ is smooth at $\beta_0$ if the vector $(\Re \delta \beta_+, \Im\delta \beta_+)$ becomes parallel to $(\Re \delta \beta_-, \Im\delta \beta_-)$ in the limit of $|\delta\beta_\pm|\rightarrow 0$ , or more precisely, 
	\begin{eqnarray}
		\lim_{|\delta\beta_\pm|\rightarrow 0}\frac{1}{|\delta\beta_+\delta\beta_-|}\det\begin{pmatrix} \Re \delta \beta_+ & \Im \delta \beta_+\\\Re \delta \beta_- & \Im \delta \beta_- \end{pmatrix}=0.
	\end{eqnarray}
	We will denote this condition as $\delta \beta_+\mathop{//}\delta \beta_-$. Furthermore, the GBZ cusps are the points that violate $\delta \beta_+\mathop{//}\delta \beta_-$.
	
	With the definition of GBZ cusps, we would like to introduce a related and useful Lemma about ``trivial'' saddle points that are defined by: 
	\begin{definition}\label{def1} 
		For a band system with the characteristic polynomial $f(E,\beta)=\det[H(\beta)-E\mathbb{I}]$, the trivial saddle point $\beta_s$ and its energy $E_s$ satisfy
		$f(E_s,\beta_s)=\partial_\beta f(E_s,\beta)|_{\beta=\beta_s}=0$ but $\partial^2_\beta f(E_s,\beta)|_{\beta=\beta_s}\neq 0$. Meanwhile, except $\beta_s$ itself, no other root of $f(E_s,\beta)=0$ shares the same modulus as $\beta_s$. 
	\end{definition} 
	The Lemma states: 
	\begin{lemma}\label{lemma1}
		A trivial saddle point $\beta_s$ on the GBZ is not a GBZ cusp, and the tangent of the GBZ at $\beta_s$ is perpendicular to the vector $(\Re \beta_s,\Im \beta_s)$.
	\end{lemma} 
	
	This Lemma offers useful information on the relation between saddle points and cusps on GBZ. It implies that only ``non-trivial" saddle points can be cusps. Therefore, if we want to find a point $\beta_s$ on GBZ that is both a saddle point and a cusp, we have to carefully adjust the parameters to achieve either (i) $f(E_s,\beta_s)=\partial_\beta f(E_s,\beta)|_{\beta=\beta_s}=\partial^2_\beta f(E_s,\beta)|_{\beta=\beta_s}=0$ or (ii) $f(E_s,\beta_s)=\partial_\beta f(E_s,\beta)|_{\beta=\beta_s}$ and $f(E_s,e^{i\theta}\beta_s)=0$ being satisfied by $\theta\neq0$. It is in line with our experience collected from concrete models. The saddle point $S_2$ depicted in Fig.~\ref{fig:process}(c) is a cusp, because it is a third-order saddle point following condition (i). Similarly, for the model discussed in the main text, the coalescence condition of saddle-point energies is compatible with condition (ii), which allows saddle points to become cusps at the PT transition point. Furthermore, except when the system is precisely tuned to the PT transition point, we find that all saddle points on the GBZ are not cusps, which is important for proving the key statements presented at the beginning of this section.  
	
	The Lemma \ref{lemma1} can be proved by examining the GBZ around a trivial saddle point. In the previous section, we have learned that for a saddle point energy $E_s$ contained in the OBC spectrum, the characteristic equation $f(E_s, \beta)=0$ has a multiple root $\beta_s$, and the GBZ equation is automatically fulfilled by $\beta_s$ itself. Then, how does the GBZ equation hold for the nearby OBC energy $E_s+\delta E$? To answer this question, we need to figure out which roots of $f(E_s+\delta E,\beta)=0$ have the same modulus. If $E_s$ and $\beta_s$ meet all the conditions listed in the Lemma \ref{lemma1}, the two candidates should be $\beta_s+\delta\beta_\pm$, where \begin{eqnarray} \label{dbeta}\delta\beta_\pm=\pm i \sqrt{\delta E\left(\frac{2\partial_E f(E,\beta)|_{E=E_s,\beta=\beta_s}}{\partial_\beta^2 f(E,\beta)|_{E=E_s,\beta=\beta_s}}\right)}\end{eqnarray} are obtained by expanding $f(E_s+\delta E,\beta_s+\delta\beta_\pm)=0$ to the leading order. Furthermore, beyond the leading order, $\delta \beta_\pm$ should be Puiseux series about $\delta E$ as \bea \label{Puiseux}\delta \beta_\pm=\sum_{k=1}^{\infty} a_{k\pm} \delta E^{k/2}.\eea The two coefficients $a_{1\pm}$ can be read from Eq.~(\ref{dbeta}) and they are related by $a_{1+}=-a_{1-}$. To satisfy the GBZ equation, $\beta_s+\delta \beta_\pm$ must have equal moduli, which means that \begin{eqnarray} |\beta_s|^2+|\delta\beta_+|^2+2\Re(\beta_s \delta\beta_+^*)=|\beta_s|^2+|\delta\beta_-|^2+2\Re(\beta_s \delta\beta_-^*).\end{eqnarray} When this equality holds, $\beta_s+\delta\beta_\pm$ are contained in GBZ. According to Eq.~(\ref{Puiseux}), the terms $|\delta\beta_\pm|^2$ and $\Re(\beta_s \delta\beta_\pm^*)$ in the above equality follow the expansions: \begin{align}
		|\delta\beta_\pm|^2&=b_{1\pm} \delta E+O(\delta E^{3/2}), \\
		\Re(\beta_s \delta\beta_\pm^*)&=c_{1\pm}E^{1/2}+c_{2\pm} \delta E+O(\delta E^{3/2}).\end{align} The two pairs of coefficients $b_{1\pm}$ and $c_{1\pm}$ can be also obtained from Eq.~(\ref{dbeta}), and it is not hard to find that $b_{1+}=b_{1-}$ and $c_{1+}=-c_{1-}$. Moreover, by comparing the coefficients on the two sides of the above equality, we can find $c_{1+}=c_{1-}$ and $c_{2+}=c_{2-}$. Thus, if the equality holds, we must have $c_{1+}=c_{1-}=0$, which causes  $\Re(\beta_s \delta\beta_\pm^*)$ to be proportional to $\delta E$ in the limit of $\delta E \rightarrow 0$. Meanwhile, $\Re(\beta_s\delta\beta_\pm^*)=|\beta_s||\delta\beta_\pm| \cos \theta_\pm$, where $\theta_\pm$ stand for the relative angles between the vectors $(\Re \delta \beta_\pm, \Im \delta \beta_\pm)$ and the vector $(\Re \beta_s, \Im \beta_s)$. By exploiting $|\delta\beta_\pm|\propto \delta E^{1/2}$, it can be shown that $\lim_{\delta E\rightarrow 0} \cos\theta_\pm=0$. Consequently, both vectors $(\Re \delta \beta_\pm, \Im \delta \beta_\pm)$ become perpendicular to the vector $(\Re \beta_s,\Im \beta_s)$ and parallel to each other when the two points $\beta_s+\delta\beta_\pm\in \text{GBZ}$ get infinitely closed to the saddle point $\beta_s\in\text{GBZ}$. According to the definition aforementioned, we finally prove that $\beta_s$ is not a GBZ cusp and the tangent of GBZ at this saddle point is perpendicular to the vector $(\Re \beta_s,\Im \beta_s)$, just as the Lemma \ref{lemma1} states. 
	
	
	It is also worth explaining why the above proof is not suitable for the two conditions mentioned above: (i) $f(E_s,\beta_s)=\partial_\beta f(E_s,\beta)|_{\beta=\beta_s}=\partial^2_\beta f(E_s,\beta)|_{\beta=\beta_s}=0$; (ii) $f(E_s,\beta_s)=\partial_\beta f(E_s,\beta)|_{\beta=\beta_s}$ and $f(E_s,e^{i\theta}\beta_s)=0$ is satisfied by $\theta\neq0$. Under condition (i), the order of the saddle point $\beta_s$ is greater than 2. There are more than two $\delta \beta$'s obtained from expanding $f(E_s+\delta E,\beta_s+\delta \beta)=0$. Nevertheless, we only need to choose two of them to fulfill the GBZ equation, e.g. $|\beta_s+\delta\beta|=|\beta_s+\delta\beta'|$. This equal modulus condition is usually not sufficient to ensure $\delta\beta\mathop{//} \delta\beta'.$ Therefore, it is permissible for the saddle points of k-th order with $k>2$ also being cusps. Under condition (ii), we can solve two roots $\delta \beta_\pm$ from $f(E_s+\delta E,\beta_s+\delta \beta)=0$. Moreover, around $\beta'=\beta_s e^{i\theta}$, we can find another $\delta \beta$ from $f(E_s+\delta E,\beta' +\delta \beta)=0$. It offers the possibility that the GBZ equation may take the form $|\beta_s+\delta\beta_+|=|\beta' +\delta\beta|$. If so, only $\beta_s+\delta\beta_+$ belongs to GBZ, while $\beta_s+\delta\beta_-$ does not. Consequently, the relations between $\delta \beta_+$ and $\delta \beta_-$ can not help us decide whether $\beta_s$ is a GBZ cusp or not. 
	
	The Lemma \ref{lemma1} has an interesting application. In Section I, we have proved that the ends of the OBC bulk spectrum always correspond to saddle points. In most scenarios, the corresponding saddle points, such as $S_{1,2}$ shown in Figs.~\ref{fig:process}(a)(b) and $S_{1,2,3}$ shown in Figs.~\ref{fig:process}(e)(f), are exactly the trivial ones defined in the Definition \ref{def1}. As a result, the behaviors of GBZ around such points can be readily inferred from the Lemma \ref{lemma1}. For a given trivial saddle point $\beta_s$ with OBC energy $E_s$, the nearby $\beta=\beta_s+\delta \beta $ on GBZ and its energy $E=E_s+\delta E$ follow the relation $\delta\beta \propto \delta E^{1/2}$. The Lemma \ref{lemma1} implies that $\Re(\beta_s\delta \beta^*)\propto \delta E$ for sufficiently small $\delta E$. Thus, the leading-order expansion of $|\beta|$ follows \begin{eqnarray} \label{beta expansion} |\beta|=\sqrt{|\beta_s|^2+|\delta \beta|^2+2\Re(\beta_s\delta \beta^*)}=|\beta_s|+ q\delta E+ \ldots,\end{eqnarray} where $q$ denotes an unimportant coefficient. The Eq.~(\ref{beta expansion}) reveals that $d |\beta|/dE$ approaches a finite value in the limit of $\delta E\rightarrow 0$. Note that $E=E_s+\delta E$ should remain in the OBC spectrum when taking the limit $\delta E\rightarrow 0$. In the main text, by investigating the frequency dependence of end-to-end Green’s functions, we have already observed the finite derivative $d |\beta|/dE$ at the trivial saddle points that stay at the band edges.
	
	We are now prepared to prove the statements presented at the beginning of this section. Since the two statements are contrapositive to each other, we simply focus on the proof of the first one. In the non-Bloch PT-exact phase, the OBC spectrum is entirely real. Therefore, to prove statement (i), we need to find the connection between the realness of the OBC spectrum and the smoothness of GBZ. Concretely, we take three pairs of points from the OBC spectrum and GBZ as $(E_0, \beta_0)$ and $(E_0+\delta E_\pm, \beta_0+\delta\beta_\pm)$ [see Fig.~\ref{fig:sketch}]. They are restricted by $f(E_0,\beta_0)=f(E_0+\delta E_+,\beta_0+\delta \beta_+)=f(E_0+\delta E_-,\beta_0+\delta \beta_-)=0$. Moreover, to use the former definition of GBZ cusps, we require that $\beta_0+\delta\beta_\pm$ are located on two sides of $\beta_0$. When the increments $\delta E_\pm$ and $\delta \beta_\pm$ become infinitely small, they are usually connected through \begin{eqnarray} \delta E_{\pm}\partial_E f(E,\beta)|_{E=E_0,\beta=\beta_0}+\delta\beta_{\pm}\partial_\beta f(E,\beta)|_{E=E_0,\beta=\beta_0} =0.\end{eqnarray} Thus, if $\beta_0\in\text{GBZ}$ is not a saddle point, namely $\partial_\beta f(E,\beta)|_{E=E_0,\beta=\beta_0} \neq 0$, the nearby $\beta_0+\delta \beta_\pm\in \text{GBZ}$ should obey \begin{eqnarray} \label{beta-E}\delta \beta_\pm = -\delta E_\pm\left(\frac{\partial_E f(E,\beta)}{\partial_\beta f(E,\beta)}\right)_{(E_0, \beta_0)} .\end{eqnarray} The real OBC spectrum results in real $\delta E_\pm $, which ensures $\delta \beta_+\mathop{//}\delta \beta_-$. As a result, GBZ is smooth at $\beta_0$. Apparently, these derivations are invalid when $\beta_0$ is a saddle point with $\partial_\beta f(E,\beta)|_{E=E_0,\beta=\beta_0} =0 $.  To make a completely smooth GBZ, it is necessary to require that the saddle points on the GBZ are not GBZ cusps. Fortunately, according to the Lemma \ref{lemma1} we just proved, this requirement is automatically fulfilled unless the parameters are fine-tuned for achieving non-trivial saddle points. The fine-tuned parameters are excluded in the proof of the statement (i), since we are interested in the most general behavior of GBZ in the non-Bloch PT-exact phase. By combining the relation Eq.~(\ref{beta-E}) and the Lemma \ref{lemma1}, we finally prove that GBZ is completely smooth in the non-Bloch PT-exact phase.
	
	
	Furthermore, it is necessary to mention the subtle difference between the ``non-Bloch PT-exact phase" and the statement that ``the OBC spectrum is completely real". As illustrated in the main text and in the  following Section IV, a system with an entirely real OBC spectrum is possibly located at the PT transition point with several saddle points on GBZ being also cusps. However, this critical situation is unstable, and it can easily break down by perturbing the parameters. For example, the model Eq.~(\ref{eq:model}) has a real OBC spectrum and also a non-smooth GBZ at the transition point $\gamma=\gamma_c$. However, the OBC spectrum becomes complex if we slightly increase $\gamma$, and the cusp on GBZ disappears if we slightly decrease $\gamma$. This critical situation is considered in the statement (ii).
	
	
	At last, we would like to emphasize the deep connection between the two key statements presented at the beginning of this section and the conformality of analytic functions. Let us review the textbook explanation of the conformality of analytic functions. We consider two smooth curves $C_1: z_1(t),t\in[0,1]$ and $C_2: z_2(t),t\in[0,1]$ on the complex plane. The two curves intersect at $z_0$, and their tangents at $z=z_0$ have a relative angle $\alpha=\arg[dz_1(t)/dt]-\arg[dz_2(t)/dt]$. After the map $w=f(z)$, we will obtain two new curves $C_1': w_1(t)=f(z_1(t)),t\in[0,1]$ and $C_2': w_2(t)=f(z_2(t)),t\in[0,1]$. The two new curves intersect at $w_0=f(z_0)$, and their tangents at $w_0$ give another relative angle $\alpha'=\arg[dw_1(t)/dt]-\arg[dw_2(t)/dt]$. When the map $w=f(z)$ is analytic and $(\partial_zf(z))_{z=z_0}\neq 0$, the conformality of analytic functions leads to $\alpha'=\alpha$. This is due to the fact that the map $w=f(z)$ only rotates the two tangents at $z_0$ by the same angle. We can easily verify this claim from $dw_{1,2}=(\partial_zf(z))_{z=z_0} dz_{1,2}$. This relation is very similar to the Eq.~(\ref{beta-E}) exploited in the proof of our key statements. Meanwhile, whether the derivatives in Eq.~(\ref{beta-E}) are well-defined is closely related to the analytic properties of the characteristic polynomial $f(E,\beta)$. For the tight-binding model with short-range hoppings, which is the focus of this work, the characteristic polynomial generally takes the form of $f(E,\beta)=\sum_{i=-M}^{N} a_i(E) \beta^i$, where $a_i(E)$ are polynomials of $E$ and $M,N$ are finite integers decided by the hopping range. $f(E,\beta)$ is singular at $\beta=0$. However, since GBZ usually bypasses the origin $\beta=0$ (except for very special cases such as taking $\gamma=t_1$ in Eq.~(\ref{HN model})), the derivatives of $f(E,\beta)$ are always well defined for $\beta$ belonging to GBZ. On the contrary, this expectation might be broken by long-range hoppings, since they possibly make $f(E,\beta)$ possess more singularities. Thus, for models with long-range hoppings, it would be interesting to further verify the two statements presented at the beginning of this section and explore their PT transition mechanism.

	\begin{figure}
		\centering
		\includegraphics[width=14cm]{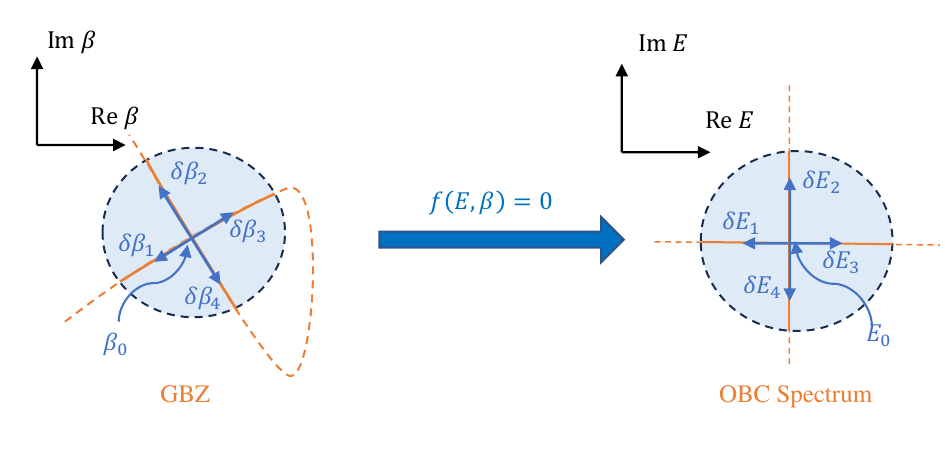}
		\caption{In single-band systems, the self-crossing GBZ induces a crossing in the OBC spectrum.}
		\label{fig:self_crossing}
	\end{figure}
	
	The conformality of analytic functions provides an intuitive way to understand the relationship between non-saddle-point GBZ cusps and the bifurcation points of complex OBC spectrums. This conformal property can further elucidate the consequence of GBZ self-intersections where the tangents are non-unique (see $\beta_0$ in Fig. \ref{fig:self_crossing}).  Similar to the cusp singularities, the GBZ self-intersection in single-band systems inevitably results in a complex OBC spectrum that belongs to the PT-broken phase, as depicted in Fig. \ref{fig:self_crossing}. For multi-band systems, GBZ is partitioned into distinct sub-GBZs. Each sub-GBZ lives on a Riemann sheet of the Riemann surface associated with the solutions of $f(E,\beta)=0$ and corresponds to different bands of the OBC spectrum \cite{Yang2019Auxiliary,hu2023multiband}. Thus, the self-intersections of GBZ in multi-bands should be separated into two types: the intersections of different sub-GBZs and the self-intersections within one sub-GBZ. Due to the energy difference between two non-Bloch bands, the intersecting sub-GBZs are actually separated from each other on the Riemann surface. Consequently, the first type of multi-band GBZ intersection does not generate  bifurcation point in the OBC spectrum (The GBZ crossing in Fig. \ref{critical} belongs to this type). In contrast, similar to the single-band cases, the self-intersection within one sub-GBZ still leads to the bifurcation in the corresponding non-Bloch band, meaning that the OBC spectrum is in the PT-broken phase. We note that all these results are in line with the cusp mechanism and do not alter its generality.

	\section{Explicit method for calculating polynomial resultant}
	
	The theory of polynomial resultant is key to our method of determining the non-Bloch PT breaking threshold. In this section, we introduce a method to effectively calculate the resultant both in theory and with the computer, with the help of the Sylvester matrix.
	
	The resultant, concerning two univariate polynomials, is a polynomial expression of their coefficients. It vanishes if and only if the two polynomials share a common root. It is commonly denoted as $\operatorname{Res}_x \left[ f(x), g(x) \right]$. We give a minimal example here: Let $f(x) = x-a$ and $g(x) = x-b$, then they have a common root if and only if $\operatorname{Res}_x \left[ f(x), g(x) \right] = a-b = 0$.
	
	A widely used explicit form of the resultant is the determinant of the Sylvester matrix \cite{lang2002algebra} 
	(see also ``Resultant $--$ Wikipedia''). Given the coefficients $f(x)=a_n x^n+\ldots+a_0$ and $g(x)=b_m x^m+\ldots b_0$, the Sylvester matrix of dimension $m+n$ is defined to be
	\begin{equation}
		\operatorname{Syl}(f,g)=
		\begin{pmatrix}
			a_n &a_{n-1}& a_{n-2} &\ldots & 0 &0& 0 \\
			0&a_n & a_{n-1} &\ldots & 0 &0& 0 \\
			\vdots&\vdots&\vdots&\ & \vdots &\vdots& \vdots  \\
			0&0&0 &\ldots & a_1&a_0& 0 \\
			0&0&0 &\ldots & a_2&a_1& a_0\\
			b_m &b_{m-1}& b_{m-2} &\ldots & 0 &0& 0 \\
			0&b_m & b_{m-1} &\ldots & 0 &0& 0 \\
			\vdots&\vdots&\vdots&\ & \vdots &\vdots& \vdots  \\
			0&0&0 &\ldots & b_1&b_0& 0 \\
			0&0&0 &\ldots & b_2&b_1& b_0\\
		\end{pmatrix}   .
	\end{equation}
	Then it is proved that
	\begin{equation}
		\operatorname{Res}_x \left[ f(x), g(x) \right] = \det \operatorname{Syl} (f,g)  .\label{resultant-syl}
	\end{equation}
	
	With the knowledge of the relation between the resultant and Sylvester matrix, we can analytically apply our method to find the non-Bloch PT breaking threshold of the Hatano-Nelson model, whose non-Bloch Hamiltonian reads \begin{eqnarray} H(\beta)=(t_1+\gamma)\beta+(t_1-\gamma)\beta^{-1}.\label{HN model}\end{eqnarray} For simplicity, we assume $t_1,\gamma>0$ below. 
	
	Independently, the threshold of this simple model can also be obtained by explicitly diagonalizing its real-space Hamiltonian $H$ under OBC. $H$ is a tri-diagonal matrix. It is straightforward to verify that the equation $H\psi_n=E_n\psi_n$ under OBC is solved by \begin{eqnarray}\psi_n(x)=\left (\frac{t_1-\gamma}{t_1+\gamma}\right)^{x/2}\sin \frac {n\pi x}{L+1}\end{eqnarray} and \begin{eqnarray} E_n = 2\sqrt{t_1^2-\gamma^2}\cos \frac {n\pi}{L+1},\label{HN energy}\end{eqnarray} where $L$ is the dimension of $H$ and $n=1,2,\dots,L$. The exponential factor in $\psi_n(x)$ is the fingerprint of NHSE. The expression of $\psi_n(x)$ also implies that GBZ is a circle on the complex plane with radius $|\beta|=\sqrt{|(t_1-\gamma)/(t_1+\gamma)|}$. The OBC eigenenergies $E_n$ exhibit a real-to-complex transition at $\gamma=t_1$, i.e., the non-Bloch PT symmetry is broken when $\gamma>t_1$.
	
	We will show how to predict the threshold $\gamma=t_1$ using our algebraic method. Following the routine given in the main text, we start by computing two resultants $g(E)= \operatorname{Res}_\beta \left[\tilde{f}(E,\beta),\partial_\beta\tilde{f}(E,\beta) \right]$ and $\operatorname{Res}_E \left[g(E),\partial_E g(E) \right]$. The polynomial $\tilde{f}(E,\beta)$ for the Hatano-Nelson model is given by $\tilde{f}(E,\beta)=\beta[f(\beta)-E]=(t_1+\gamma)\beta^2-E\beta+(t_1-\gamma)$. By applying Eq.~(\ref{resultant-syl}), \begin{eqnarray} g(E)=\det \begin{pmatrix} t_1+\gamma & -E & t_1-\gamma \\ 2t_1+2\gamma& -E & 0\\ 0& 2t_1+2\gamma& -E \end{pmatrix}=(t_1+\gamma)[4(t_1^2-\gamma^2)-E^2].\end{eqnarray} Before computing the second resultant, we pause to elaborate the significance of $g(E)$. The roots of $g(E)=0$ are $E_\pm=\pm2\sqrt{t_1^2-\gamma^2}$, which are exactly the energies of the two saddle points $\beta_\pm=\pm\sqrt{(t_1-\gamma)/(t_1+\gamma)}$ that satisfy $\partial_\beta H(\beta)=0$. Similarly, the saddle-point energies of a general model can also be obtained from the solution of $g(E)=0$. Then we compute the resultant $\operatorname{Res}_E \left[g(E),\partial_E g(E) \right]$. By inserting $g(E)=(t_1+\gamma)[4(t_1^2-\gamma^2)-E^2]$, we derive that \begin{eqnarray} \operatorname{Res}_E \left[g(E),\partial_E g(E) \right]=-(t_1+\gamma)^3\det \begin{pmatrix} 1 & 0 & -4(t_1^2-\gamma^2) \\ 2& 0 & 0\\ 0& 2& 0\end{pmatrix}=16(t_1+\gamma)^3(t_1^2-\gamma^2).\end{eqnarray} In the main text, we have claimed that the parameters satisfying $\operatorname{Res}_E \left[g(E),\partial_E g(E) \right]=0$ should contribute to saddle-point energy degeneracy. This can be readily checked here. $\operatorname{Res}_E \left[g(E),\partial_E g(E) \right]=0$ has a solution $\gamma=t_1$, and the saddle-point energies $E_\pm$ indeed become degenerate at $\gamma=t_1$.
	
	Since the model Eq.~(\ref{HN model}) is very simple, the solution $\gamma=t_1$ of $\operatorname{Res}_E \left[g(E),\partial_E g(E) \right]=0$ is already enough to predict its PT breaking threshold. Nevertheless, for illustrative purposes, we will continue with the remaining procedure of our method for this model. Generally, we need to insert the roots of $\operatorname{Res}_E \left[g(E),\partial_E g(E) \right]=0$ back into $g(E)=0$ to find the degenerate energy $E_s$, and then keep those roots corresponding to a real degenerate $E_s$ that is given by the saddle points contained in GBZ. For this model, $g(E)=-2\gamma E^2$ when $\gamma=t_1$. Thus, $E_s=0$ is a degenerate root of $g(E)=0$, which is also the energy of the saddle point $\beta_s=0$ satisfying $\partial_\beta H(\beta)=0$. On the other hand, the GBZ of this model contracts to the origin $\beta=0$ when $\gamma=t_1$. Since $\gamma=t_1$ satisfies the criteria aforementioned as $\Im E_s=0$ and $\beta_s\in \text{GBZ}$, we ascertain that it should be the PT breaking threshold. The prediction of our algebraic approach is consistent with the previous result obtained below Eq.~(\ref{HN energy}). 
	

	As a final remark, the Sylvester determinant is not always the fastest way to obtain the resultant, and calculating it by hand can be a formidable task. Fortunately, there are several alternative methods as well as symbolic computer programs that help us do this. Function names like \texttt{resultant(f(x), g(x), x)} are available in many software programs. For instance, we used MATLAB to obtain the numerical results in the main text.
	
	\section{DOS near non-Bloch van Hove singularity}
	
	We first show how to explicitly derive the formal expression of the DOS for a real bulk spectrum. Its form resembles the familiar one in the Hermitian case:
	\begin{eqnarray}
		\rho(E)=\frac{1}{2\pi}\sum_{\beta(E)\in {\rm GBZ}} \left|{\rm Im} \left[\frac{1}{\beta\partial_\beta H(\beta)}\right]_{\beta=\beta(E)}\right|.
	\end{eqnarray}
	
	The linear DOS (states per unit length) is defined to be $\rho(E)=\lim_{L\to\infty} L^{-1}\sum_{n=1}^L \delta(E-E_n)$, where $E_n$'s are energy eigenvalues. When the spectrum is purely real and $E\in\mathbb{R}$, we are allowed to write the DOS in another form. Mathematically, the $\delta$-function can be written as $\lim_{\epsilon\to0^+}\frac{1}{E-E_n-i\epsilon}=i\pi\delta(E-E_n)+\mathcal{P}(\frac{1}{E-E_n})$, where $\mathcal{P}$ is the principal value. To translate the expression $\rho(E)=\lim_{L\to\infty}L^{-1}\sum_{n=1}^L\delta(E-E_n)$ into the imaginary part of Green's function, the limit $L\to\infty$ controls the number of these $\delta$ functions, while the limit $\epsilon\to0^+$ controls the width of the spectra line shape at $E_n$ that mimics a $\delta$ function. Due to their independent roles in the calculation of the DOS, the two limits are exchangeable, giving rise to
	\begin{eqnarray}
		\rho(E) = \lim_{\epsilon\to 0^+}\lim_{L\to\infty} \frac{1}{\pi L} \Im \sum_{n=1}^{L}\frac{1}{E-E_n-i\epsilon}.
	\end{eqnarray}
	To proceed, we rewrite the right-hand side in terms of the Green's function. With a spectral decomposition, the Green's function can be written as
	\begin{equation}
		G(E) =\sum_{n=1}^L \frac{\ket{nR}\bra{nL}}{E-E_n} ,
	\end{equation}
	where $\ket{nR}$ and $\bra{nL}$ are right and left eigenstates of $H$, respectively. Then we find the relation 
	\begin{eqnarray}
		\rho(E) = \lim_{\epsilon\to 0^{+}}\lim_{L\to\infty} \frac{1}{\pi L} \Im \operatorname{tr} \left[G(E-i\epsilon)\right].
	\end{eqnarray}
	
	When the system size $L$ approaches infinity, thanks to the well-established theory on non-Hermitian Green's function \cite{xue2021simple}, we are able to re-express $\rho(E)$ as a contour integral along the GBZ
	\begin{eqnarray}
		\rho(E)= \lim_{\epsilon\rightarrow 0^+} \frac{1}{\pi }{\rm Im} \oint_{{\rm GBZ}} \frac{d \beta}{2\pi i\beta} \frac{1}{E-H(\beta)-i\epsilon} \label{dos},
	\end{eqnarray}
	where $H(\beta)=\sum_{n=-l}^{r} h_n \beta^n$ is the non-Bloch Hamiltonian. The integral in Eq.~(\ref{dos}) can be directly calculated through the residue theorem. It has been proved that there are always $l$ poles surrounded by the GBZ, when $\epsilon$ takes a finite positive value \cite{Okuma2020, Zhang2020correspondence}. So these $l$ poles $\beta_1, \dots, \beta_l$ (sorted by their moduli) contribute $l$ residues. The residue at the pole $\beta_i$ is $R_i(E-i\epsilon)=-\left(\beta\partial_\beta H(\beta)\right)^{-1}_{\beta=\beta_i}$. Straightforwardly, the DOS can be obtained by summing the imaginary parts of the residues at $\beta_1,...,\beta_l$ and taking the limit $\epsilon\to 0^+$, so we have
	\begin{eqnarray}
		\rho(E)=\frac{1}{\pi}{\rm Im} \sum_{i=1}^l R_i(E-i0^+)
	\end{eqnarray}
	and also
	\begin{eqnarray}
		\rho(E)=-\frac{1}{\pi}{\rm Im} \sum_{i=l+1}^{r+l} R_i(E-i0^+)
	\end{eqnarray}
	because $\sum_{i=1}^{r+l} R_i(E-i0^+)=0$. Meanwhile, by reversing the sign of the infinitesimal parameter $\epsilon$, one can find two alternative expressions:
	\begin{eqnarray}
		\rho(E)=-\frac{1}{\pi}{\rm Im} \sum_{i=1}^l R_i(E-i0^-),\\
		\rho(E)=\frac{1}{\pi}{\rm Im} \sum_{i=l+1}^{r+l} R_i(E-i0^-).
	\end{eqnarray}
	An important observation connecting the two sets of expressions is that the two poles $\beta_l$ and $\beta_{l+1}$ will switch their order by shifting $\epsilon$ from $0^+$ to $0^-$, since their moduli at $\epsilon=0$ are forced to be equal by the GBZ equation $|\beta_l|=|\beta_{l+1}|$. Accordingly, the residues are related by
	\begin{eqnarray}
		R_l(E-i0^+)=R_{l+1}(E-i0^-),\ R_{l+1}(E-i0^+)=R_{l}(E-i0^-).
	\end{eqnarray}
	Using this relation and that the two summations $\sum_{i=1}^{l-1} R_i(E-i\epsilon)$ and $\sum_{i=l+2}^{r+l} R_i(E-i\epsilon)$ are continuous at $\epsilon=0$, it is easy to prove that
	\begin{eqnarray}
		\rho(E)=\frac{{\rm Im} R_{l}(E-i0^+)-{\rm Im} R_{l+1}(E-i0^+)}{2\pi}=-\frac{{\rm Im} R_{l}(E-i0^-)-{\rm Im} R_{l+1}(E-i0^-)}{2\pi}.\label{dos0}
	\end{eqnarray}
	Next, we will show ${\rm Im} R_{l}(E-i0^+)>0>{\rm Im} R_{l+1}(E-i0^+)$ which guarantees the DOS within the spectrum to be positive and therefore physical. Through differentiating both sides of $H(|\beta|e^{i\theta})=E-i\epsilon$ like
	\begin{eqnarray}
		\left[{\rm Re}\left(\beta\partial_\beta H(\beta)\right)+i{\rm Im}\left(\beta\partial_\beta H(\beta)\right)\right](d \log|\beta|+id \theta)=dE-id \epsilon,
	\end{eqnarray}
	we find 
	\begin{eqnarray}
		\frac{d \log |\beta_i(E-i\epsilon)|}{d\epsilon}={\rm Im} \left[\frac{1}{\beta\partial_\beta H(\beta)}\right]_{\beta=\beta_i}=-{\rm Im} R_i(E-i\epsilon)\label{relation}
	\end{eqnarray}
	Since an infinitesimal positive $\epsilon$ moves the two poles $\beta_l$ and $\beta_{l+1}$ (originally both being on the GBZ) inward and outward the GBZ, respectively, two compatibility conditions
	\begin{eqnarray}
		\left(\frac{d \log |\beta_l(E-i\epsilon)|}{d\epsilon}\right)_{\epsilon=0^+}<0,\\ 
		\left(\frac{d \log |\beta_{l+1}(E-i\epsilon)|}{d\epsilon} \right)_{\epsilon=0^+}>0
	\end{eqnarray}
	must hold. These conditions lead to ${\rm Im} R_{l}(E-i0^+)>0>{\rm Im} R_{l+1}(E-i0^+)$ via Eq.~(\ref{relation}). Finally, taking the absolute values of ${\rm Im} R_{l}(E-i0^+)$ and ${\rm Im} R_{l+1}(E-i0^+)$ in Eq.~(\ref{dos0}), we reach
	\begin{eqnarray}
		\rho(E)=\frac{1}{2\pi}\sum_{\beta(E)\in {\rm GBZ}} \left|{\rm Im} \left[\frac{1}{\beta\partial_\beta H(\beta)}\right]_{\beta=\beta(E)}\right|.\label{dos1}
	\end{eqnarray}
	This is the formula for the DOS used in the main text.
	
	Next, we move on to the behavior of DOS near a spectral singularity. Inserting $\partial_\beta H(\beta_s) = 0$, we immediately see that a saddle point accompanies a divergence in DOS. Besides, it is also of great interest to know the asymptotic behavior of DOS near a non-Bloch van Hove singularity, which will need us to expand $H(\beta)$ to the leading order near a saddle point.
	
	We have defined a saddle point to be the solution to both $H(\beta)-E = 0$ and $\partial_\beta\left[H(\beta)-E \right] = 0$, which is equivalent to saying $\beta$ is a multiple root to $H(\beta)=E$. Naturally, we can further characterize a saddle point by its multiplicity. In other words, a $k$-th order saddle point is a root to $H(\beta)-E$ of multiplicity $k$. Near a $k$-th order saddle point $\beta=\beta_s$, we expand $H(\beta)$ into
	\begin{equation}    \label{eq:expansion}
		H(\beta) \approx H(\beta_s) + \left.\dfrac{\partial_{\beta}^{(k)} H(\beta)}{k!}\right|_{\beta=\beta_s} (\beta-\beta_s)^k= H(\beta_s)+ A(\beta-\beta_s)^k    ,
	\end{equation}
	where we have replaced the derivative by a coefficient $A \neq 0$ for simplicity. So the leading order of $\partial_\beta H(\beta)$ is
	\begin{equation}    \label{eq:derivative}
		\partial_\beta H(\beta) \approx kA (\beta-\beta_s)^{k-1}    .
	\end{equation}
	
	Alternatively, we are able to solve $\beta$ as an inverse function. Let $\delta E = E_{\rm{bulk}}-E_s = H(\beta)-H(\beta_s)$, we have
	\begin{equation}    \label{eq:betas}
		\beta \approx \beta_s+\left(\frac{\delta E}{A}\right)^{\frac{1}{k}} e^{i\frac{2\pi n}{k}} \quad n=0,1,\dots,k-1 .
	\end{equation}
	Note carefully, although there are in total $k$ branches of solutions, only a part of them belong to the GBZ. Nevertheless, all branches lead to qualitatively the same contribution to the DOS, so we are safe to keep the factor $e^{i\frac{2\pi n}{k}}$ for now. Inserting Eq.~(\ref{eq:betas}) back into Eq.~(\ref{eq:derivative}), we get
	\begin{equation}
		\partial_\beta H(\beta)\approx kA^{\frac{1}{k}}(E-E_s)^{1-\frac{1}{k}}e^{-i\frac{2\pi n}{k}}    .
	\end{equation}
	Thus,
	\begin{equation}
		\beta\partial_\beta H(\beta)\approx k\beta_sA^{\frac{1}{k}}(E-E_s)^{1-\frac{1}{k}}e^{-i\frac{2\pi n}{k}}+k (E-E_s)  .
	\end{equation}
	The second term $\delta E$ is much smaller than the first term $(\delta E)^{1-1/k}$ in the limit $\delta E\to 0$, so we can drop the second term. Finally, we have come to the conclusion that the DOS near a non-Bloch van Hove singularity scales as
	\begin{equation}    \label{eq:asym}
		\rho(E) \propto |E-E_s|^{\frac{1}{k}-1} .
	\end{equation}
	Writing this as $\rho(E) \propto |E-E_s|^{-\alpha}$, we find the exponent to be
	\begin{equation}
		\alpha = 1-\dfrac{1}{k} .
	\end{equation}
	
	\section{A different type of saddle point coalescence}
	
	\begin{figure}[h]
		\centering
		\includegraphics[width=16cm]{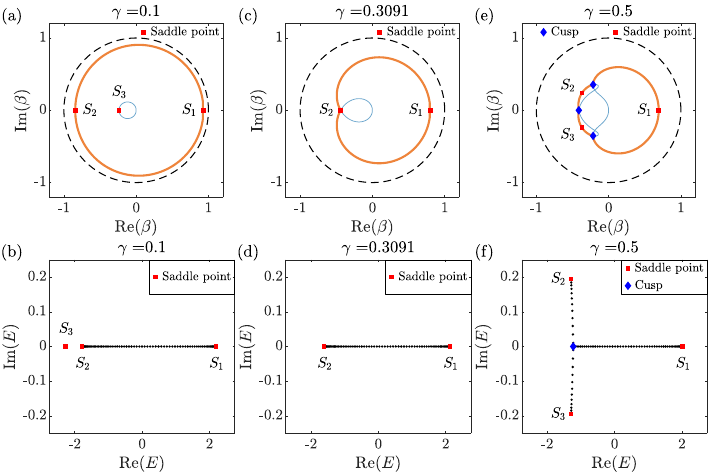}
		\caption{\label{fig:process}The flow of saddle points through the non-Bloch PT breaking threshold in model Eq.~(\ref{eq:model}) with $t_1=1$ and $t_2=0.1$. The threshold parameter is $\gamma_c=0.3091$. The system size is fixed to $L=80$. Saddle points are denoted by $S_n$. Another saddle point $S_4$ is not shown in these figures since it is far from the GBZ and irrelevant to PT breaking. }
	\end{figure}
	
	\begin{figure}[h]
		\centering
		\includegraphics[width=12 cm]{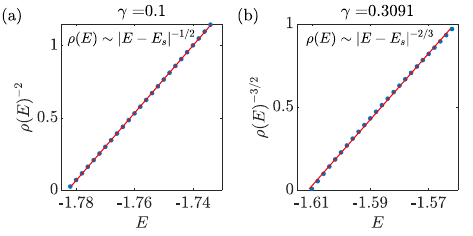}
		\caption{\label{fig:fit}Linear regression of the DOS near the left band edge. $E_s$ denotes the energy of the left edge, which is also a saddle point. (a) PT symmetric phase, with $\gamma=0.1$ and $E_s=-1.783$. (b) Critical point of PT breaking, with $\gamma=0.3091$ and $E_s=-1.610$. }
	\end{figure}
	
	\begin{figure}[h]
		\centering
		\includegraphics[width=10cm]{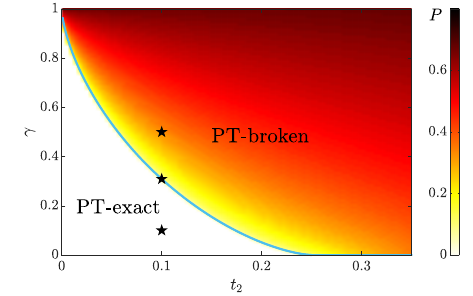}
		\caption{\label{fig:phase}The non-Bloch PT phase diagram of the model $H(\beta)=(t_1+\gamma)\beta+(t_1-\gamma)\beta^{-1}+t_2(\beta^2+\beta^{-2})$ with $t_1=1$.}
	\end{figure}
	
	The process of non-Bloch PT breaking discussed in the main text is about two second-order saddle points coalescing at one common energy. Besides, a different type of coalescence is found in a PT symmetric model with non-Bloch Hamiltonian
	\begin{equation}    \label{eq:model}
		H(\beta)=(t_1+\gamma)\beta+(t_1-\gamma)\beta^{-1}+t_2(\beta^2+\beta^{-2}) .
	\end{equation}
	
	From Fig.~\ref{fig:process}, we find that this non-Bloch PT breaking falls into the same pattern described in the main text, that the symmetry breaking is accompanied by coalescing of two saddle points. However, the process here is more unique than the example in the main text, reflected in two aspects.
	\begin{enumerate}
		\item The coalescence is between $S_2$ and $S_3$. In the PT symmetric phase, only $S_3$ is outside the spectrum, which differs from the case in the main text.
		\item At PT breaking transition point, two saddle points not only coalesce at a common energy but also share the same (complex) momentum on the GBZ ($S_2$ in Fig.~\ref{fig:process}(c)).
	\end{enumerate}
	Also, two saddle points merging into one naturally leads to a third-order saddle point, i.e.,
	\begin{equation}
		H(\beta)-E
		= \partial_\beta H(\beta)
		= \partial_\beta^2 H(\beta)
		= 0
	\end{equation}
	holds at $S_2$.

	It may seem that this PT breaking transition does not induce any significant physical change, since no new singularity with divergent DOS has been created. But taking a closer look we find a drastic change in the asymptotic behavior near the left end of the band. This change originates from the fact that two second-order saddle points merge to become a third-order saddle point. According to Eq.~(\ref{eq:asym}), the local behavior of the DOS correspondingly changes from $\left| E-E_s \right|^{-1/2}$ to $\left| E-E_s \right|^{-2/3}$. This subtle difference has been verified by numerical linear regression, shown in Fig.~\ref{fig:fit}.

	At last, we remark that the algebraic method for determining the PT breaking threshold still works well within the current model Eq.~(\ref{eq:model}). We demonstrate a PT phase diagram from numerics in Fig.~\ref{fig:phase}.

	\section{Typical multi-band transitions}
	
	The non-Bloch PT symmetry breaking is ubiquitous in models with arbitrary number of bands, as we claimed in the main text. To complement the argument in the main text and above in these Supplemental Materials, we investigate here a typical multi-band model. We shall see that the PT symmetry can break at cusp singularities on GBZ expectedly, or at a band exceptional point. The latter is directly related to the multi-band nature of the model, since in the parameter space of momentum at least two sheets are required to support an EP. 
	
	We study the non-Hermitian Su-Schrieffer-Heeger (SSH) model, in which the NHSE was first appreciated \cite{yao2018edge}
	\begin{eqnarray}
		H_2(\beta)=\begin{pmatrix}
			0&t_1+t_2\beta^{-1}+(t_3+\gamma)\beta \\t_1+t_2\beta+(t_3-\gamma)\beta^{-1} & 0
		\end{pmatrix}. \label{two-band model}
	\end{eqnarray}
	The non-Hermiticity is reflected in the asymmetric hopping between nearest sites. It is known that the PBC spectrum and the OBC spectrum are distinct at a generic parameter. In the continuum limit, they are written as $E_\pm(\beta)|_{\beta \in \text{GBZ}}$ and $E_\pm(\beta)|_{\beta \in \text{BZ}}$, respectively, with $E_\pm(\beta)=\pm \sqrt{Q(\beta)}$ and 
	\be
	Q(\beta)=t_1^2+t_2^2+t_3^2-\gamma^2+t_1(t_2+t_3)(\beta+\beta^{-1})+t_1\gamma(\beta-\beta^{-1})+t_2t_3(\beta^2+\beta^{-2})+t_2\gamma(\beta^2-\beta^{-2}) .
	\ee
	
	\begin{figure}[h]
		\includegraphics[width=16cm]{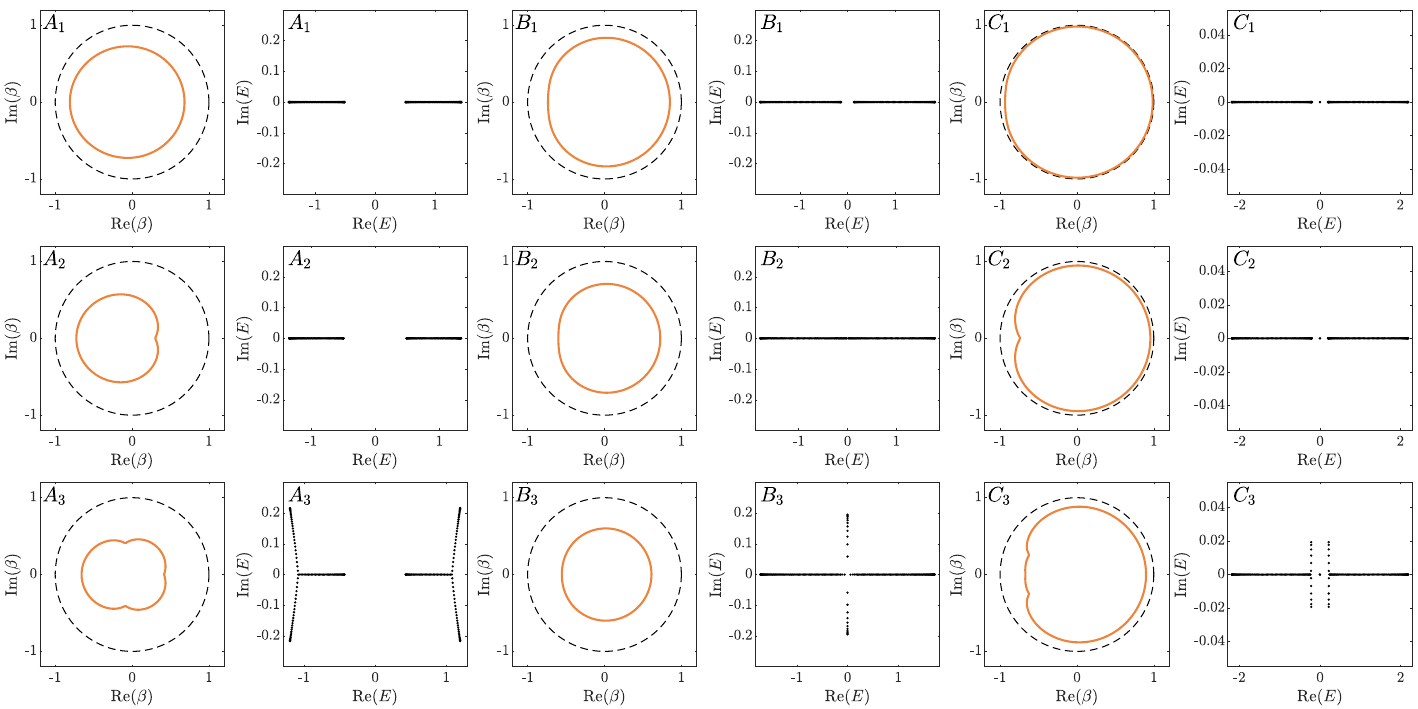}	
		\caption{GBZs and OBC spectrums during typical non-Bloch PT breaking transitions in the model $H_2(\beta)$.   The system size is $L=100$. The parameters are $t_1=1,t_2=0.3$ and $A_1$: $t_3=0.18,\gamma=0.15$, $A_2$: $t_3=0.18,\gamma=0.2556$, $A_3$: $t_3=0.18,\gamma=0.40$, $B_1$: $t_3=0.55,\gamma=0.15$, $B_2$: $t_3=0.55,\gamma=0.283$, $B_3$: $t_3=0.55,\gamma=0.40$,  $C_1$: $t_3=0.9,\gamma=0.02$, $C_2$: $t_3=0.9,\gamma=0.0648$, $C_3$: $t_3=0.9,\gamma=0.15$. Their positions are marked in the phase diagram Fig.~\ref{phase1}. }\label{typical transitions}
	\end{figure}
	
	We find that the spectrum is PT symmetric, i.e., entirely real, when $Q(\beta)$ is real and non-negative. In the PBC case, the spectrum is always complex as long as $\gamma\neq 0$. As to the OBC case, we shall find several possibilities for PT breaking. We present three groups of transitions (group A, B, and C) in Fig.~\ref{typical transitions}. Group A and Group C are quite akin to the example in Fig.~\ref{fig:process}. The spectrum ramifies at an end (but at opposite ends, comparing A and C), indicating that $Q(\beta)$ undergoes the same process. On the other hand, Group B is saliently different. The bands are seen to close the gap at the origin and then grow into the imaginary direction, indicating $Q(\beta)$ preserves real, but a part of it changes sign. This is a new type of transition we have not encountered in the main text. An interesting observation is that the number of saddle points of the OBC spectrum does not change before and after the transition. At $E=0$, we can ask what exactly happened to the non-Bloch Hamiltonian $H_2(\beta)$. There are two possibilities: one being $H_2(\beta) \propto \sigma_{\pm}$ so $H_2(\beta)$ is at an EP; the other being $H_2(\beta) = 0_{2\times 2}$. The latter is a rare case, and it turns out it is not the case in group B. It demands fine-tuning for the null matrix to appear. More thorough discussions on these EPs on the GBZ can be found in \cite{Yokomizo2020topological}.
	
	To conclude, we found two types of PT breaking in this non-Hermitian SSH example:
	\begin{itemize}
		\item Type I associated with the cusp singularities on GBZ (group A and C), 
		\item Type II associated with the exceptional points on the GBZ (group B). 
	\end{itemize}
	
	Although the PT breaking mechanism is multi-fold, the non-Bloch PT breaking threshold can still be determined by the very same procedure as in the main text. To be specific, we write the characteristic polynomial of the non-Hermitian SSH model as $f_2(E,\beta)=\beta^2\text{det}\left[H_2(\beta)-E\mathbb{I}_2\right]$. Then, all the saddle point energies are given by
	\begin{eqnarray}
		g_2(E)=\text{Res}_\beta\left[f_2(E,\beta),\partial_\beta f_2(E,\beta)\right]=0
	\end{eqnarray}
	and different saddle points meet when
	\begin{eqnarray}
		\text{Res}_E\left[g_2(E),\partial_\beta g_2(E)\right]=0.
	\end{eqnarray}
	
	\begin{figure}[h] 
		\centering
		\includegraphics[width=10cm]{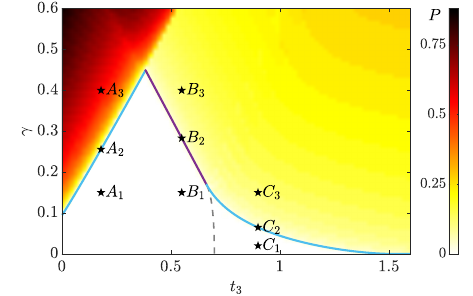}
		\caption{The non-Bloch PT phase diagram of the model $H_2(\beta)$ in Eq.~(\ref{two-band model}) with $t_1=1,t_2=0.3$.}\label{phase1}
	\end{figure}

	Among all solutions to this equation, what concern us are only the ones contained in the GBZ. The valid solutions compose the lines demonstrated in Fig.~\ref{phase1}. The solid line precisely matches the PT phase boundary in Fig.~\ref{phase1}. The lighter (darker) color stands for the transition belonging to type I (type II). Exceptionally, the dashed line that is also a part of the solutions does not indicate any PT transitions. We look carefully into this line, and find that here neither GBZ cusps nor exceptional points are generated. Thus, the only possibility is that $H_2(\beta)$ is a null matrix at the gap closing point $Q(\beta)=0$, exactly the fine-tuned case we skipped before. The existence of this dashed line shows that the saddle point coalescence is only necessary but not sufficient for the PT breaking in this two-band model. Nevertheless, the saddle point coalescence condition still works well in searching for PT breaking thresholds. This is reminiscent of searching for PT transition in Bloch bands using the band gap closing condition, which is also only a necessary condition. 
	
	\begin{figure}[h]
		\centering
		\includegraphics[width=16cm]{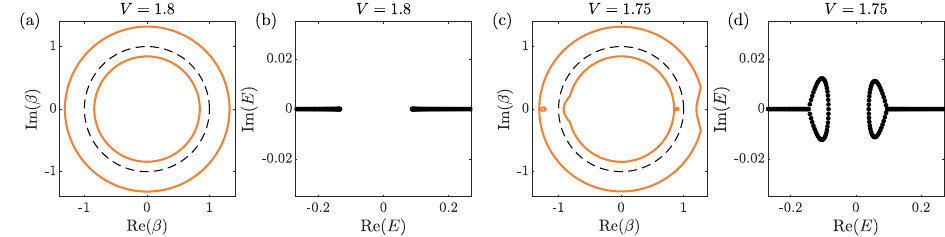}
		\caption{The OBC spetrums and GBZs of the model $H_c(\beta)$ with $t_1=t_2=\delta=1$, $\gamma_1=0.2$, $\gamma_2=-0.3$ and (a)(b): $V=1.8$, (c)(d): $V=1.75$. The spectrums are diagonalized with system size $L=600$.} \label{critical}
	\end{figure}
	
	Beyond this specific model, general multi-band non-Bloch PT transitions should also fall into the two categories introduced above. Let us consider a slightly different multi-band transition in the model
	\begin{eqnarray}
		H_c(\beta)=\begin{pmatrix}
			(t_1-\gamma_1)\beta^{-1}+(t_1+\gamma_1)\beta+V&\delta\\\delta &(t_2-\gamma_2)\beta^{-1}+(t_2+\gamma_2)\beta-V
		\end{pmatrix}.
	\end{eqnarray}
	For this model, the non-Bloch PT symmetry is exact in Fig.~\ref{critical}(b) while is broken in Fig.~\ref{critical}(d). Unlike all the examples discussed above, the OBC spectrum in the PT broken phase possesses ``ring structures", and the energy point with the maximum imaginary part is no longer a saddle point. Nevertheless, the PT transition still changes the number of saddle points contained in the OBC spectrum, which decreases from $4$ in Fig.~\ref{critical}(b) to $2$ in Fig.~\ref{critical}(d). Accompanied by this process, GBZ cusps appear in Fig.~\ref{critical}(c), and this transition belongs to category I.

\end{document}